\documentclass[12pt,a4paper]{article}
\usepackage[latin1]{inputenc}
\usepackage{amsmath}
\usepackage{amsfonts}
\usepackage{amssymb}
\title{Multi Phase Analytical description Of Vaporization And Condensation In Nanofluid Droplets}
\author{Fredrick Michael*}

\begin{document}
\maketitle                                             
\begin{abstract}
 Recently there has been a growing interest in nanodroplet engineering and the dynamics of vaporization and condensation of fluids at the nanoscale. The applications of nanodroplet engineering is in the biomolecular field, physical chemistry, and engineering power production and storage devices technology.... other applications are in aerosols, fuels and turbulence in power source design. In this paper a theoretical model for multiphase nanofluids is derived from information theory and temperature dependent distributions are obtained for the Lennard-Jones potential interactions of the nanofluid droplets.

\end{abstract}

The simple paradigm of a nanofluid is an inert material such as Argon, as a fluid at low temperatures, in a droplet of a diameter a few nano meters in scale. Such a nanodroplet contains from the tens to the hundreds of Argon atoms in a fluid phase. The kinematics of the fluid nanodroplet as it is accelerated and it interacts with other materials, gas, liquid, molecules or solid state materials is the dynamics we wish to address here. 

   A simple example for what we will model by theory here is the nanodroplet of Argon, say containing a few hundred Argon atoms, moving at a uniform velocity towards a non chemically reactive solid surface that could represent in more complex experiments or applications a reactive or non reactive surface of a device for power production such as in fuel cell walls, a porous membrane in biological cellular reactions, or reactive chemistry analysis such as lab-on-a-chip. The medium here is a vacuum though more complex arrangements could be of secondary gases or liquids.

   At the nanoscale the aggregation of atoms and the screening effect that occurs as the number of atoms increases towards a size of the micron range is less pronounced , and quantum mechanical effects tend to emerge and be more pronounced. This is a general phenomenon and is noted in large molecules as opposed to assemblages of molecular chains, in nanoparticles as opposed to bulk solids. Examples are the magnetic hysteresis in Gold and Palladium nanoparticles that are diamagnetic in Bulk \cite{fred1} . Another bulk equivalently large scale effect are the continuum bands of electronic transport and the breakdown of these bands and with the subsequent localization of electronic orbitals described by hopping Hamiltonian methods.
 
 An observed effect at the nanoscale of the breakdown of continuum electronic bands is the quenching of plasmon oscillations. For a discussion of these quantum nano scale effects see any recent book on nanoparticle engineering.  Electronic Plasma is a Fermi fluid description of the collective bulk effect of interacting electrons in continuum bands. This has been used to very good design advantage in absorption and emission spectrum of interaction with photons and the electronic plasma in image enhancement and medical applications. 
 
  The emergent quantum phenomenon in nanofluid droplets is expected to be similarly novel and of interest in engineering and applications. There is then great interest in accurate models of multiphase processes and the dynamics of nanofluids in general and of nanodroplets in particular. Multiphase processes of phase transitions are of importance as discussed in vaporization applications such as in modeling vaporization in turbulent regimes and fuel mixing, in biomolecular applications, etc.  
   
   Theoretically the modelling of nanoscale fluids should be done at the fully quantum mechanical level. This means the interacting many-particle physics approach, many particle Hamiltonians and diagrammatic methods. However the screening effect of aggregate atoms allows for approximate descriptions at various length nano and mesoscopic scales where either fully quantum mechanical effects emerge or partially screened effects are noted and can be described by semi classical approximations at the longer length scales from  the nanometer $10^{-9} m$ to $10^{-6} m$ micrometer range. At larger length scales it is expected that classical descriptions of fluids are accurate for 'bulk' fluid description.  
   
   \section{Fully Quantum Mechanical Theory}
A fully quantum mechanical theory of nanofluid droplets has already been written and is published in the many particle theory of Fermi systems phase transitions. The review articles by Mattuck and Johanssen are of very influential importance in this theoretical approach \cite{mattuck1, mattuck2}. The theory is a description of a fluid and a vapor phase coexisting and described simultaneously in a matrix propagator or Green's function. The quantum mechanical Hamiltonian is a many-particle Hamiltonian of a form that can be written in occupation number formalism or in second quantization as sums of creation $c^\dagger$ and annihilation $c$ operators of the observables of kinetic and potential energies. 
\begin{equation}
\hat{H}=  \sum \limits_{\vec k} ^{}  \epsilon_{\vec{k}} {c^\dagger}_{\vec{k}}{c}_{\vec{k}} + \sum \limits_{\vec k,\vec{l},\vec{m},\vec{n}}{} V_{\vec{k},\vec{l},\vec{m},\vec{n}} {c^\dagger}_{\vec{k}}{c}_{\vec{l}}{c^\dagger}_{\vec{m}}{c}_{\vec{n}}.    \label{eq1} 
\end{equation}
The multiphases of interest are solid, liquid and vapor or gaseous phases. These are described by the inclusion of translation periodic vectors for the solid phase such that the Hamiltonian of Eq.(\ref{eq1}) becomes periodic with respect to the lattice structure of the solid phase,the lattice periodicity vector we set to be a constant $\vec{Q_o},\vec{Q' _o}$ for simplicity. The Hamiltonian of the solid phase becomes 

\begin{equation}  
\hat{H}= \sum \limits_{\vec{k},\vec{Q}}^{\inf} \epsilon_{\vec{K}+\vec{Q_{o}}}{c^\dagger}_{\vec{k}+\vec{Q_o}}
{c}_{\vec{k} } 
 + \sum \limits_{\vec k,\vec{l},\vec{m}, \vec{n}, \vec{Q_o},\vec{Q' _o}}^{\inf} V_{\vec{k}+\vec{Q_o},\vec{l},\vec{m}+\vec{{Q'}_o},\vec{n}} {c^\dagger}_{\vec{k}+\vec{Q_o}} {c}_{\vec{l}}{c^\dagger}_{\vec{m}+\vec{Q'_o}}{c}_{\vec{n}}. \label{eq2}.
\end{equation}
As the temperature is increased, the solid melts to a fluid phase, and the Hamiltonian is described by a similar Hamiltonian, however the lattice symmetry or periodicity is broken, and the lattice vector which was set to be a constant $\vec{Q_o}$ for simplicity as in Eq.(\ref{eq2}) is now a nonperiodic momentum vector and the fluid is described by a Hamiltonian of the form of Eq.(\ref{eq1}). Note that the solid is described by an effective mass electronic Hamiltonian describing band electronic effective mass. This effective mass simple description is no longer applicable in the fluid phase and actual electronic mass or effective atomic charge mass and effective nuclei is to be used in the Hamiltonian approximation, or a full interacting Hamiltonian of the protons and the electrons with description of the interactions of the protons and the electrons in each atom of the fluid and between atoms....This would mean adding in superposition to Eq.(\ref{eq1}) terms of the  potential interaction between protons and terms of potential interaction between protons and electrons such that
\begin{equation}
<KE_p>+ <V_{p-p}>=
\sum\limits_{\vec {o}}^{} {\epsilon^p}_{\vec{o}} {c^\dagger}_{\vec{o}}{c}_{\vec{o}}
 +
  \sum\limits_{\vec{o},\vec{p},\vec{q},\vec{r}} ^{} {V^{p-p}}_{\vec{o},\vec{p},\vec{q},\vec{r}} {c^\dagger}_{\vec{o}} {c}_{\vec{p}}{c^\dagger}_{\vec{q}} {c}_{\vec{r}} 
\end{equation}
   and interaction terms between the protons and electrons of the similar form $<V_{p-e}>= \sum \limits_{\vec {o},\vec{p},\vec{q},\vec{r}}^{} {V^{p-e}}_{\vec{o},\vec{p},\vec{q},\vec{r}} {c^\dagger}_{\vec{o}}{c}_{\vec{p}}{c^\dagger}_{\vec{q}}{c}_{\vec{r}} $, though the choice of potential can differ for effective charge interactions in band material, for effective Hydrogen like atoms, etc.
 
 The gaseous or vapor phase is again described by a similar Hamiltonian as in Eq.(\ref{eq1}), the difference here being that again as in the fluid phase Hamiltonian we can no longer use the approximations we used in the solid phase of effective mass band electrons. Also, the Hydrogen like atom with effective charge and effective nuclei that is useful for simple atoms description in the fluid phase is applicable for simple atoms.   Otherwise we would again have to fully describe each electron and proton in each atom in the fully interacting picture we have discussed, with electron and proton and proton-electron interaction potentials in superposition with the electronic Hamiltonian. 
 The finite temperature and the number of the atoms in each phase are the important considerations we discuss next. The number of atoms in each phase is dictated by the chemical potential; the number of particles is the summation of the Fermi distribution function as in $<N>=\sum \limits_{k}{} \frac{1}{e^{\beta(\epsilon_k-\mu)} + 1}$. How one partitions the groups of electrons , protons, and atoms they comprise depends on the level of approximation used as discussed.
 
  The Hamiltonians can be written as a matrix Hamiltonian, or as one single Hamiltonian using spinors. The matrix Hamiltonian also means that the Green's functions of the Hamiltonians can be written as a matrix Green's function. The Hamiltonians would then be written as
\begin{eqnarray}
<\hat{H}>=  
 \left[ 
 \begin{matrix}
{<\hat{H}_{solid-Solid}>}{<\hat{H}_{solid-Liquid}>}{<\hat{H}_{Solid-Gas/vapor}>} \\
{<\hat{H}_{Liquid-Solid}>}{<\hat{H}_{Liquid-Liquid}>}{<\hat{H}_{Liquid-Gas/vapor}>}\\
{<\hat{H}_{Gas/vapor-Solid}>}{<\hat{H}_{Gas/vapor-Liquid}>}{<\hat{H}_{Gas/vapor-Gas/vapor}>} \label{matrixHamiltonian1}
\end{matrix} \right]. 
\end{eqnarray}   

        The Green's function in quantum mechanics is simply the response function, and it can be described as the inverse of the Hamiltonian as 
        $G^{r}(\vec{k},\omega)=\frac {-i}{\hbar \omega -\upsilon_{\vec{k}} + i\eta} $, here the superscript $G^{r}$ indicates the retarded Green's function and the imaginary infinitesimal $+i\eta$ in the denominator insures no divergence in the Green's function by shifting the argument into the complex plane and the total kinetic and self energy is written in shorthand as $\upsilon_{\vec{k}}=\epsilon_{\vec{k}}- i\Sigma({\vec{k},\omega})$ \cite{Olle1}. Physically the causality of incoming or outgoing waves is represented by this shift into the complex plane.
        
             The Green's function for the total Hamiltonian of the coexisting phases is then also a matrix, and can be written as in Eq.(\ref{matrixHamiltonian1}), \begin{eqnarray}
   G= \left[  \begin{matrix}
     {g_{s-s}}\;\;\;{g_{s-L}}\;\;\;{g_{s-v}}\\{g_{L-s}}\;\;\;{g_{L-L}}\;\;\;{g_{L-v}}\\{g_{v-s}}\;\;\;{g_{v-L}}\;\;\;{g_{v-v}}
     \end{matrix} \right] , \label{matrixGreen'sfunction1}
\end{eqnarray}
   and here the Green's functions are subscripted in simple notation such as the solid-vapor interacting Green's function $g_{s-v}$ and so on for the solid-solid , liquid-liquid and the vapor-vapor (gaseous) phases. 
  
   The observables of interest are the total number of particles, the number of particles or atoms in a particular phase, the rate of vaporization or of condensation or solidification, temperature and energy dependences of the phase transitions. These are obtained in expectation values of the Green's functions, also to be considered as probability functions analogues here, in summation with the observable of interest. The total number of Argon atoms for example, $<N>=<N_{s-s}>+<N_{L-L}>+<N_{v-v}>$ is obtained by the diagonal trace of the Green's function matrix since $<N>=\int G(\vec{k}, \omega)\frac{d\vec{k}}{(2\Pi)^3} \frac{d\omega}{2\Pi}$, and the trace consists of the individual phases such as for example $<N_{s-s}>=\int {g_{s-s}}(\vec{k}, \omega)\frac{d\vec{k}}{(2\Pi)^3}\frac{d\omega}{2\Pi}$. Fourier transforming from the energy to the time domain and taking a time derivative can describe vaporization from the liquid phase for example $\frac{\partial<N_{L-v}>}{\partial t}= \frac{\partial}{\partial t}\int g_{L-v}(\vec{k},t)\frac{d\vec{k}}{(2\Pi)^3}dt$, while a similar transform and time derivative can describe sublimation from the solid to the vapor phase $\frac{\partial<N_{s-v}>}{\partial t}= \frac{\partial}{\partial t}\int g_{s-v}(\vec{k},t)\frac{d\vec{k}}{(2\Pi)^3}dt$.
   
     The solution of the equations for the observables, as they are rather complicated mathematical expressions, can be done by computer calculations of approximate summation methods such as the Hartree-Fock approximation, mean field and self consistent methods and dynamic mean field and other similar summation methods, and only rarely in the formal diagrammatic theory are closed form solutions possible, as in linearized mean field like approximations. other approaches to analytic solutions are possible by transformation methods, and these are discussed in later work.

\section{Semi Classical Theory. Lennard Jones Fluids.} 
  The nanofluid droplets have been described reasonably successfully in the literature by semi classical methods \cite{semiclassical1,semiclassical2}. The  approaches have been to treat the nanofluid droplet by hydrodynamic means, the vapor by Boltzmann equation, and to evolve the number of particles and obtain rates of change in number of particles. Approximations are made such as Boltzmann equation \cite{semiclassical2} collision terms of a certain order, an example is the force collision terms up to a third order expansion by \cite{russianauthor}. 
  
  In hydrodynamics approaches, full analytic solutions are not obtained to the hydrodynamic equations... approximations of linearization and similar assumptions for solution are made such as in \cite{semiclassical1} , examples of approximations are of viscosity and compressibility.

  Another approach is possible. Given the success of recent molecular dynamics simulations in describing experimental results, approximating the nanofluid as a Lennard-Jones fluid seems reasonable. Then a similar theory can be made at the Lennard-Jones approximation level that is applicable in the range of applicability of the Boltzmann and hydrodynamics equations' scale of applicability. That is from the hundreds of atoms to the micrometer scale. From an analogy to the fully quantum mechanical phase transition theory discussed in the previous section, a theory of coexisting phases of nanofluid droplets can be obtained. The advantages would be the simultaneous description of co-existing phases by analogy to the matrix formulation discussed, long range order parameter approach as in the diagonal trace and summation to obtain particle or number of atoms in the individual phases.

    A theoretical advantage of the full quantum theory is the Hamiltonian superposition that allows us to include external perturbations and interaction terms. From the point of view of the approach we will pursue in order to derive a Lennard-Jones fluids coexisting phases theory, the superposition principle is replaced by the Legendre transform of the maximum entropy method of theory derivation, and equivalently the information theoretic method. Alternatively a Lennard-Jones semi classical Lagrangian or Hamiltonian may be used, and dynamics with or without a heat bath and therefore Fokker-Planck equations obtained in matrix form. We proceed from the information theoretic and equivalently maximum entropy derived theory approach.  
  
\subsection{Derivation}  
   The entropy measure to be used we choose to be the Gibbs-Boltzmann entropy, $<S>=-c \int{P(\vec{r},t)lnP(\vec{r},t)}{d\vec{r}}$ which can be generalized to the nonextensive entropy of C. Tsallis \cite{tsallis1,fred2} for better accuracy and other novel effects such as the q-parametrization of the deviation of the PDF from the Gaussian to the power-law PDF. Note that nonlinearity due to interactions naturally or physically deviate the PDF from the Gaussian distribution towards the power-law distribution. The nonextensive statistics has been noted for theoretically producing this form of power-law statistics from a formally similar mathematics to the traditional thermodynamics and information theory. The connections are more profound and review articles and applications can be found at \cite{tsallis1} .   
  
   The maximum entropy approach is then of the maximization of the entropy measure constrained with observables of the system in a Legendre transform with Lagrange multipliers setting the weighted units to the entropic measure's.

   The choice of observables is one of choosing the minimum set of observables that describe the physics of interest. These observables are  described mathematically as functions of the other observables chosen as representative of the observable dynamics of the system. A discussion of this was made by the original author of maximum entropy theory \cite{jaynes1, jaynes2} E. T. Jaynes.  
 
  We choose observables analogous to the quantum mechanical observables in the fully quantum mechanical theory. The momentum or kinetic energy observables of quantum mechanics correspond to the first and second moments of the variables of position, the potential energy or interaction potential terms are similarly included here though we choose the Lennard-Jones potential and not the Coulomb potential, and the addition of velocity variables insures that a connection to classical equations of motion such as hydrodynamic and Boltzmann is made.
  The observable moments are:
\begin{eqnarray}  
 M^2 (\Delta\vec x_{ij})= \sum \limits_{i,j}^{N_v}    <(\Delta \vec x_{ij} - <\Delta\vec x_{ij}>)^2> \nonumber \\ =\int (\Delta\vec x_{ij} - <\Delta\vec x_{ij}>)^2 P(\Delta\vec{x_{ij}},\Delta\vec{v_{ij}},\Delta{t}) {d\Delta\vec{x_{ij}}} {d\Delta\vec{v_{ij}}}   
  \end{eqnarray}
  \begin{eqnarray}  
 M^2 (\Delta\vec v_{ij})= \sum \limits_{i,j}^{N_v}    <(\Delta \vec v_{ij} - <\Delta\vec v_{ij}>)^2> \nonumber \\ =\int (\Delta\vec v_{ij} - <\Delta\vec v_{ij}>)^2 P(\Delta\vec x_{ij},\Delta\vec v_{ij},\Delta{t}) d\Delta\vec x_{ij} d\Delta\vec v_{ij}   
  \end{eqnarray}
 
 \begin{eqnarray}
 <V_{vapor-vapor}>= \sum \limits_{i,j}^{N_v}  < ( { \frac{\sigma_{vv}} {{r_{ij}}^6} }  - { \frac{\eta_{vv}} { r_{ij}^{12}} } )> \nonumber \\
 = \sum \limits_{i,j}^{N_v}  \int ( { \frac{\sigma_{vv}} {{r_{ij}}^6} }  - { \frac{\eta_{vv}} { r_{ij}^{12}} } ) P(\Delta\vec x_{ij},\Delta\vec v_{ij},\Delta t) {d\Delta\vec{ x_{ij}}}{ d\Delta\vec{v_{ij}}} \end{eqnarray}
\pagebreak
  Here we have chosen to use the Lennard-Jones potential in full form as appropriate to the level of mechanics and scale. Note that in solid form a harmonic oscillator approximation has been used by some researchers and in fluid phase this is somewhat questionable. The liquid and the vapor phases interactions are chosen to be described by Lennard-Jones potentials. The Lennard-Jones potential is a scale dependent approximation for screened interactions of coulomb like potential and include the repulsive term where no two atoms can occupy the same space. The potential behaves as an inverse $\frac{a}{r^{b}}$ Coulomb-like potential at far enough distances. In quantum mechanics the repulsion of atoms at almost vanishing distances is the Fermi exclusion principle statistics or for zero or integer spin atoms and particles the Bose statistics which as known can overlap and condense. As discussed in the previous sections the potential in the quantum mechanical interaction theory is a coulomb or screened coulomb potential. 
  \pagebreak
  
     The maximization of the entropy measure with the observables constraining the maximization allows us to derive a least biased distribution \cite{jaynes1}. As an example we derive the vapor phase least biased distribution, 
      \begin{equation}
  \delta <S>+\delta [\beta{M^2 (\Delta\vec x_{ij})} + \alpha {M^2 (\Delta\vec x_{ij})} + \gamma <V_{v-v}>  ]=0.
  \end{equation}
  The least biased distribution is
 \begin{equation} 
P(\Delta\vec x_{ij},\Delta\vec v_{ij},\Delta t)=\frac{ 
e^{-\beta (\Delta\vec x_{ij} - <\Delta\vec x_{ij}>)^2 
-\alpha (\Delta\vec v_{ij} - <\Delta\vec v_{ij}>)^2 - 
\gamma ( \frac{\sigma_{vv}}{{r_{ij}}^6}   -  \frac{\eta_{vv}}{ r_{ij}^{12}}   ) } } { Z(\Delta t)}  
\end{equation} 
   The Lagrange multipliers can be evaluated from the identity
\begin{equation}
\frac{1}{Z(t)} \frac{\partial {Z(t)}} {\partial{\beta}}= 
-<(\Delta\vec x_{ij} - <\Delta\vec x_{ij}>)^2>,   \label{lagrangemultiplier1}
\end{equation}
similar relations between observables and for the other Lagrange multipliers can be evaluated.
 
  The partition function is the inverse of the normalization of the probability distribution function and as usual is the sum over the distribution and its variables as in $Z(t)=\int P(x,y,..,v_{z},t)dxdy..d{v_{z}}$.

     The matrix of co existing phases here is obtained similarly to the fully quantum mechanical theory. The method of matrix or spinor Hamiltonian and derivation of Green's functions is discussed in the quantum mechanical case and will not be revisited here. we proceed to the discussion of the vapor and liquid phases and the interaction terms.
  
  Due to the use of similar Lennard-Jones interaction potentials between the liquid-liquid atoms, the liquid-vapor atoms, and the vapor-vapor atoms and the solid-solid atoms, the mathematical distinction between the vapor and liquid phases is one of number of atoms in an individual phase and not the analytic form of the least biased distribution function of particles in a liquid solid or vapor phase. This is made clear when we note that the nanofluid droplet of number of atoms $<N_L>$, where $<N>=<N_L>+<N_v>+<N_s>$ is a constant number and is obtained by summing the distribution of atoms over the variables of the  observables 
  
  \begin{eqnarray} 
 <N_L> = \sum_{ij}^{N_L} \int P(\Delta\vec x_{ij},\Delta\vec v_{ij},\Delta t) {d\Delta\vec x_{ij} }d{\Delta\vec v_{ij}} \\
  <N_v> =\sum_{ij}^{N_v} \int   P(\Delta\vec x_{ij},\Delta\vec v_{ij},\Delta t) {d\Delta\vec x_{ij} }d{\Delta\vec v_{ij}}, \\
  \vdots  \nonumber \\
  \vdots \nonumber \\
 <N_{vs}>= \sum_{ij}^{N_v,N_s} \int   P(\Delta\vec x_{ij},\Delta\vec v_{ij},\Delta t) {d\Delta\vec x_{ij} }d{\Delta\vec v_{ij}}, \\
  \vdots \nonumber \\
  \vdots
\end{eqnarray}  

 The use of the Lennard-Jones potential in both the vapor and fluid phases is the cause of the formal similarity of the terms here. If we had chosen to use the harmonic oscillator potential for the solid phase interactions, for one reason or another, and the Lennard-Jones potential for the vapor and a similar screened Coulomb potential for the liquid phase, or had chosen to treat the solid phase as fully quantum mechanical while the liquid phase was treated as a Lennard-Jones liquid while say the vapor phase was treated also by a Lennard-Jones or by a Boltzmann equation PDF, the mathematical analytic similarity would no longer exist, and the offdiagonal terms in the matrix of the Argon atoms distribution functions would no longer seem redundant. The matrix of these distribution functions is in direct analogy to the fully quantum mechanical theory
 
  \begin{eqnarray}
   P= \left[  \begin{matrix}
     {p_{s-s}}\;\;\;{p_{s-L}}\;\;\;{p_{s-v}}\\{p_{L-s}}\;\;\;{p_{L-L}}\;\;\;{p_{L-v}}\\{p_{v-s}}\;\;\;{p_{v-L}}\;\;\;{p_{v-v}}
     \end{matrix} \right] , \label{matrixPDF'sfunction1}
\end{eqnarray}
   
    Other instances where offdiagonal terms and therefore offdiagonal long range order gives no information is the quantum mechanical mathematical analogue of the mean field Ferromagnetic model, where the preferred direction and the linearization of the field terms makes the offdiagonal terms redundant. As such they are constant terms in the original Hamiltonian superposition and are zero terms in the  Green's function matrix analogous to the PDF matrix here, which then contains only diagonal terms and only diagonal long range order. 
 
   Here then the analogy to the diagonal long range order distributions (re: Green's functions) in the quantum mechanical phase transition theory has the diagonal PDFs as the distributions importance in determining the observables of interest such as the total particle number and inter-relationship of the phases in determining rates of change of one phase's number of particles or atoms as dependent upon the other phases' number of atoms or particles. The redundant seeming offdiagonal terms are also similarly able to provide the information about the rates of change of number of atoms from one phase to another, and this is discussed subsequently. Also to be considered is the symmetry between offdiagonal terms such as $p_{s-L}<=?=>p_{L-S}$.     
   
      The similarity of potential terms used in all the phases, here the Lennard-Jones potential used in all three co-existing phases makes the coexisting vapor and fluid and solid terms of individual phases described by the diagonal terms, the analogues of the diagonal long range order in the quantum mechanical phase transition theory, similar in form to the offdiagonal terms and denotes the importance of the number of atoms or particles in each phase as the distinguishing feature in our simplified approach. We note again that the use of dissimilar potentials or even statistics such as quantum mechanical descriptions of one phase and semi-classical descriptions of another phase, would result in distinct differences in analytical form of the distributions and the perceived redundancy of offdiagonal terms would no longer exist, and it is possible that the importance then of offdiagonal long range order would outweigh the diagonal terms. Note that we use the traditional terminology of diagonal and offdiagonal long range order in analogy to the fully quantum mechanical phase transition theory.

         The similarity of potential terms used in all the phases, here the Lennard-Jones potential used in all three co-existing phases makes the coexisting vapor and fluid and solid terms of individual phases described by the diagonal terms, the analogues of the diagonal long range order in the quantum mechanical phase transition theory, similar in form to the offdiagonal terms and denotes the importance of the number of atoms or particles in each phase as the distinguishing feature in our simplified approach. We note again that the use of dissimilar potentials or even statistics such as quantum mechanical descriptions of one phase and semi-classical descriptions of another phase, would result in distinct differences in analytical form of the distributions and the perceived redundancy of offdiagonal terms would no longer exist, and it is possible that the importance then of offdiagonal long range order would outweigh the diagonal terms. Note that we use the traditional terminology of diagonal and offdiagonal long range order in analogy to the fully quantum mechanical phase transition theory. 
   
   The total number of particles is a constant in our simple 100 atom Argon nanodroplet as pointed out, and is the sum described as $<N>=<N_L>+<N_v>+<N_s>$. The number of liquid particles at a time $t$ is the $<N_L>=<N>-<N_v>-<N_s>$, and the rate of change of this number is the partial time derivative $\frac{\partial <N_L>}{\partial t}=-\frac{\partial <N_v>}{\partial t}-\frac{\partial <N_s>}{\partial t}$. This corresponds to the time rate of change in the number of liquid atoms that are being vaporized or solidified. Other interaction distributions, say between the liquid and vapor atoms $<N_{v-L}>=\sum \limits_{ij}^{N_v,N_L}\int  P(\Delta\vec x_{ij},\Delta\vec v_{ij},\Delta t) {d\Delta\vec x_{ij} }d{\Delta\vec v_{ij}}$ are useful in rates of transition, where a time derivative is the rate at which liquid Argon droplet atoms vaporize to the gaseous phase.
   These offdiagonal terms are important in offdiagonal long range order derivation, and here in rates of transition from one phase to another. They are the semi classical analogues of the quantum mechanical Green's functions of interacting vapor-liquid and liquid-vapor and solid-liquid and solid-vapor Hamiltonians. Differences are that the quantum mechanical off diagonal Green's functions are Hermitian transposes of each other. Here the analogue offdiagonal PDFs are to be considered from the point of view of reversibility and mathematical similarity. Transpose of coordinates in our two-point PDF functions does not result in a change of sign for de-meaned second moments, however it is possible that with moments about the mean, the mean or drift terms could introduce a sign change upon interchange of coordinates. Care must be take in any assumptions about the symmetry of offdiagonal terms. The even power form of the Lennard-Jones potential implies that interchanging the interaction coordinates does not cause a change in sign overall however this might not be the case for dissimilar potentials and statistics between phases as mentioned. 
   
\subsection{Derivation of Solutions}
  An interesting point to note is that we can obtain closed form solutions for the $<N_L>$ and $<N_v>$ and $<N_s>$ PDFs . The PDF for the nanofluid droplet is  
       
\begin{equation}
  P(\Delta\vec x_{ij},\Delta\vec v_{ij},\Delta t)= \frac{ e^{-\beta (\Delta\vec x_{ij} - <\Delta\vec x_{ij}>)^2 -\alpha (\Delta\vec v_{ij} - <\Delta\vec v_{ij}>)^2 - \gamma ( \frac{\sigma_{vv}} {{r_{ij}}^6}  - \frac{\eta_{vv}}{ r_{ij}^{12}} ) } } { Z(\Delta t)} . 
\end{equation}      

   The Lagrange multipliers are evaluated as discussed from the relationship between partition function and the observables $\frac{\partial lnZ}{\partial \beta}=-M^2 (\Delta\vec x_{ij})$ and similarly for the other observables. Note that the Lagrange multipliers of the second moments are related to the diffusion coefficients by the temperature-diffusion coefficient thermodynamics relation $D=\frac{1}{2\beta}$, and $<(x_{ij})^2)>=2Dt$. 
   The analytic solution of interest here is the one of the transformation of the Lennard-Jones potential to a drift coefficient term. This is done by uncompleting the square of the Gaussian
   \begin{eqnarray}
    p(X)= exp{  (X^2 - 2 X<X> + <X>^2)/2Dt}
    \end{eqnarray}
  and comparing it with the $r_{ij}$ dependent terms of the de-mean (or constant or linear and renormalized mean) Gaussian

     \begin{equation} 
    P(r_{ij},....)= e^{-\beta(  (r_{ij}^2 - 2 r_{ij} <r_{ij}> + <r_{ij}>^2 )- \gamma LJ(r_{ij})  )   }e^{.....}e^{.....}.....
   \end{equation}
   The Lennard-Jones potential terms can be seen to be terms that can be transformed to a drift coefficient or the first moment by un-completing the square. We obtain 
   \begin{equation}
   < r_{ij}>={r_{ij}}  + \sqrt{ {r_{ij} }^2    +  \frac{\gamma}{\beta} LJ(r_{ij}) }.  \label{LJdrift1}
   \end{equation}
   
 \subsection{Analytic distribution functions}  
   With this the $r_{ij}$ dependent terms become a simple $e^{-\beta (r_{ij}-<r_{ij}>)^2 }$ Gaussian which by the identity  $\frac{\partial lnZ}{\partial \beta}=-M^2 (\Delta\vec x_{ij})$ gives the simple diffusion-temperature relation $\beta=\frac {1} {2 D_{r} t}$ with the temperature from thermodynamics $\beta=\frac{1}{k_{B} T}$.
   
     The resulting PDF is a Gaussian with the coordinates expressed in spherical coordinates for simplicity. The integration for the partition function or normalization can be accomplished, however we use the relationship $\beta Z(t)^2=2\pi $ and the resulting distribution is
   \begin{equation}
  P(r_{ij},....)=  \frac {     e^{  \frac{-  (\vec r_{ij} -<\vec r_{ij}> )^2}  {2{D_r}t} }  
   }  {  \sqrt{4\pi {D_r} t} ^3} \frac {     e^{  \frac{-  ({\vec v^r}_{ij} -<{\vec v^r}_{ij}> )^2}  {2{D_{v^r}}t} }  
   }  {  \sqrt{4\pi {D_{v^r}} t} ^3 }.
  \end{equation} 
 Note that aside from the $r_{ij}$ terms which have the complicated Lennard-Jones potential dependence and its radial coordinates dependence Eq.(\ref{LJdrift1}) added to the radial coordinate linear drift, the other variables such as the velocity magnitudes in different coordinates are all simply zero mean $<\theta_{ij}>=0$, constant drift $<\theta_{ij}>=b$ or linear drift terms corresponding to accelerated motion $<\theta_{ij}>= c\theta_{ij}+b$; Note that a constant drift value mean is useful in uniform motion descriptions, and a linear $<\theta_{ij}>= c\theta_{ij}+b$  drift term corresponds to a constant driving force term. 
 \section{Computation and numerical simulation.}
  The simulation of the statistical dynamics of a nanofluid droplets is accomplished by using these probability distribution functions in a Monte Carlo or equivalently Stochastic simulation. In the next of the series of papers on nanofluid droplets, the simulation of nanofluid Argon droplet is performed using a particular set of parameters. We choose to simulate a nano particle of number of Argon atoms of 100, at room temperature $300K$, in a vacuum and with no surfaces to interact with for the initial simulation of vaporization. The simple high temperature droplet parameters mean we can neglect the solid phase and the $3X3$ matrices become $2X2$ matrices with no loss of generality. The Diffusion coefficients are obtained from the inverse of the temperature and the Boltzmann constant as discussed. The Lennard-Jones parameters are obtained from CHARMM Molecular Dynamics Argon file data from experimental measured potential parameters.
   The trajectory of $<N>=100$ Ar atoms is obtained from the stochastic differential equations of the individual atoms. The initial fluid configuration is assumed to be a fixed configuration, and initially the fluid droplet is at zero coordinate. The initial configuration is assumed to be almost solid Argon which is FCC face center cubic, and the average nearest neighbour distance is 3.6 Angstroms, and the next nearest neighbour distance is obtained from that by geometrical analysis and so on. Alternatively we simulate a periodic initial configuration with variation of nearets neighbor distances about the mean nearest neighbor distance of the FCC solid phase value. This initial configuration is almost immediately melted as the high room temperature in comparison to fluid Argon temperature of approximately $67K$ causes the atoms to move apart rapidly. The movement is somewhat restricted by the potential which is attractive at long distances and repulsive at short distances, and the fluid phase begins to vaporize as atoms separate from the nano fluid droplet and move independently off. As in other literature, we define the vaporized atoms as atoms that move 3 nearest neighbor distances away from the nearest fluid atoms that are all within 1.5 times the nearest neighbor distances to each other. In simulation coordinates, we can plot the trajectories of atoms and can observe the vaporization of Argon atoms from the nanofluid droplet. The simulation is run until half of the atoms are vaporized. We also estimate this by tracking the rate of change of number of fluid atoms.

\section{Conclusion}
In this article we have derived a Lennard-Jones fluid theory of nanofluid droplet statistical dynamics from an information theory and equivalently maximum entropy approach. We have derived analogously to the fully quantum mechanical theory of phase transitions of fluids a matrix of probability distributions and therefore number of Argon atoms that are in a particular solid, fluid or vapor phase at any given time. We discuss the current approach's analogues of diagonal and off diagonal long range order in particle or atom number summation and rates of transition from one phase to another. We then solve analytically a particular distribution function and obtain analytic forms that are normalized. This analytic form is due to the radial dependence of the Lennard-Jones potential. We then discuss research of computational simulations performed on nanofluid droplets which we will report in detail on for materials of Argon for simplicity, water as a check of more complex nanofluids and for future applications to biomolecular modeling in nanofluids.

\pagebreak
 *Fredrick Michael. Michael Research R$\&$D. fmicha3@uic.edu. fnmfnm2@yahoo.com. 773-641-0894.

\end{document}